\documentclass[%
reprint,
showpacs,preprintnumbers,
amsmath,amssymb,
aps,
]{revtex4-1}
\usepackage{epsf}
\usepackage{graphicx}
\usepackage{color}
\usepackage{amsmath}
\usepackage{amsfonts}
\usepackage{amssymb}
\usepackage{dcolumn}%

\begin{document}
\title{Current-induced spin torque resonance of magnetic insulators}
\date{\today}
\author{Takahiro Chiba$^{1}$, Gerrit E. W. Bauer$^{1,2,3}$, and Saburo
Takahashi$^{1}$}
\affiliation{$^{1}$ Institute for Materials Research, Tohoku University, Sendai, Miyagi
980-8577, Japan}
\affiliation{$^{2}$ WPI-AIMR, Tohoku University, Sendai, Miyagi 980-8577, Japan}
\affiliation{$^{3}$ Kavli Institute of NanoScience, Delft University of Technology,
Lorentzweg 1, 2628 CJ Delft, The Netherlands}

\begin{abstract}
We formulate a theory of the AC spin Hall magnetoresistance (SMR) in a bilayer
system consisting of a magnetic insulator such as yttrium iron garnet (YIG)
and a heavy metal such as platinum (Pt). We derive expressions for the DC
voltage generation based on the drift-diffusion spin model and quantum
mechanical boundary condition at the interface that reveal a spin torque
ferromagnetic resonance (ST-FMR). We predict that ST-FMR experiments will
reveal valuable information on the current-induced magnetization dynamics of
magnetic insulators and AC\ spin Hall effect.

\end{abstract}
\maketitle


Ferrimagnetic insulators such as yttrium iron garnet (YIG) with high critical
temperatures and very low magnetization damping have been known for decades to
be choice materials for in optical, microwave or data storage technologies
\cite{Wu}. Near-dissipationless propagation of spin waves make YIG wires and
circuits interesting for low power data transmission and logic devices. A
crucial breakthrough was the discovery that the magnetization in YIG can be
excited electrically by Pt contacts \cite{Kajiwara}, thereby creating an
interface between electronic/spintronic and magnonic circuits. However, the
generation of coherent spin waves by the current-induced spin orbit torques in
Pt is a strongly non-linear process and the low critical threshold currents
found by the experiments \cite{Kajiwara} cannot yet be explained by theory
\cite{Xiao}. Here we suggest and model a simpler method to get to grips with
the important current-magnetization interaction in the YIG$|$Pt system without
a problematic threshold, \textit{viz}. by employing the recently discovered
magnetoresistance of YIG$|$Pt bilayers or Spin Hall Magnetoresistance (SMR)
\cite{Nakayama,Chen} to detect current-induced spin torque ferromagnetic
resonance (ST-FMR).

The SMR\ is the dependence of the electrical resistance of the normal metal on
the magnetization angle of a proximity insulator and is caused by a concerted
action of the Spin Hall Effect (SHE) \cite{review12} and its inverse (ISHE).
An alternative mechanism of the SMR phenomenology in terms of an equilibrium
proximity magnetization close to the YIG interface has been proposed
\cite{Chen}. However, this interpretation has been challenged by experiments
\cite{Nakayama,Althammer}. Moreover, while experiments of many groups are
described quantitatively well by the SMR model with one set of parameters
\cite{Vlietstra,Hahn1,VlietstraX,Isasa,Marmion}, we are not aware of a
transport theory that explains the observed magnetoresistance in terms of a
monolayer-order magnetic Pt.

Current-induced spin torque ferromagnetic resonance (ST-FMR) has been
demonstrated \cite{Liu,Kondou,Ganguly} in bilayer thin films made from
metallic ferromagnets (FM) and nonmagnetic metals (N). In these experiments
the SHE transforms an in-plane alternating current (AC) into an oscillating
transverse spin current. The resultant spin transfer resonates with the
magnetization at the FMR frequency. The effects induced simultaneously by the
Oersted field can be distinguished by a different symmetry of the resonance on
detuning. The magnetization dynamics leads to a time dependence of the bilayer
resistance by the anisotropic magnetoresistance (AMR). Mixing the applied
current and the oscillating resistance generates a DC voltage that is referred
to as spin torque diode effect \cite{Tulapurkar,Sankey}.

The longitudinal spin Seebeck effect was found to be frequency independent up
to 30 MHz \cite{Roschewsky}. The DC ISHE induced by spin pumping has been
observed by many groups, but detection of the AC spin Hall effect
\cite{Jiao13} has only recently been reported in metallic structures
\cite{Weiler0,Wei,Hyde} as well as in Pt$|$YIG under parametric microwave
excitation \cite{Hahn2}. A DC voltage can be generated in Pt$|$YIG under
FMR conditions by rectification of the AC spin Hall effect by means of the
SMR, but this signal was found to be swamped by the DC spin Hall effect
\cite{Iguchi}. A study of the spin Hall impedance concludes that the material
constants of Pt$|$YIG bilayers do not depend on frequency up to 4 GHz
\cite{Lotze}.

\begin{figure}[ptb]
\centering\includegraphics[width=0.45\textwidth,angle=0]{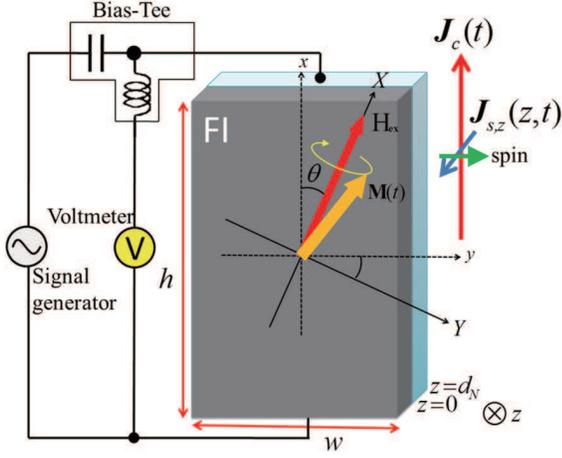}\caption{Schematic
set-up to observe the SMR mediated spin torque diode effect. The light blue
rectangle is a normal metal (N) film with a finite spin Hall angle, while F is
a ferromagnetic insulator. The F$|$N bilayer film is patterned into a strip
with width\ $w$ and length $h$. A Bias-Tee allows detection of a DC voltage
under an AC bias.}%
\label{ACSMR}%
\end{figure}

In this paper we suggest to combine the principles sketched above to realize
ST-FMR for bilayers of a ferro- or ferrimagnetic insulator (FI) such as YIG
and a normal metal with spin orbit interaction (N) such as platinum
\cite{Mosendz10} (see Fig. \ref{ACSMR}). We derive the magnetization dynamics
and DC voltages generated by the SMR-induced spin torque diode effect as a
function of the external magnetic field. Our theory should help to better
understand the elusive current-induced magnetization dynamics of ferromagnetic
insulators which should pave the way for low-power devices based on magnetic
insulators \cite{Wu}.

The spin current through an F$|$N interface is governed by the complex
spin-mixing conductance $G^{\uparrow\downarrow}$ \cite{Brataas00}. The
prediction of a large $\operatorname{Re}G^{\uparrow\downarrow}$ for interfaces
between YIG and simple metals by first-principle calculations \cite{Jia11} has
been confirmed by recent experiments \cite{Burrowes12,Weiler}. The spin
transport in N (spin Hall system) can be treated by spin-diffusion theory with
quantum mechanical boundary conditions at the interface to the insulating
ferromagnet \cite{Chen,Chiba}. The AC current with frequency $\omega_{a}=2\pi
f_{a}$ induces a spin accumulation distribution $\boldsymbol{\mu}_{s}(z,t)$ in
N that obeys the spin-diffusion equation
\begin{equation}
\partial_{t}\boldsymbol{\mu}_{s}=D\partial_{z}^{2}\boldsymbol{\mu}_{s}%
-\frac{\boldsymbol{\mu}_{s}}{\tau_{\mathrm{sf}}}, \label{tdsd}%
\end{equation}
where $D$ is the charge diffusion constant and $\tau_{\mathrm{sf}}$ spin-flip
relaxation time in N. In position-frequency space the solution for the
spatiotemporal dependence of the spin accumulation reads $\boldsymbol{\mu}%
_{s}(z,\omega)=\boldsymbol{A}e^{-\kappa(\omega)z}+\boldsymbol{B}%
e^{\kappa(\omega)z}$, where $\kappa^{2}(\omega)=(1+i\omega\tau_{\mathrm{sf}%
})/\lambda^{2}$, $\lambda=\sqrt{D\tau_{\mathrm{sf}}}$ is the spin-diffusion
length, and the constant column vectors $\boldsymbol{A}$ and $\boldsymbol{B}$
are determined by the boundary conditions for the spin current density in the
$z$-direction $\boldsymbol{J}_{s,z}(z)$, where $\boldsymbol{J}_{s,z}%
/\left\vert \boldsymbol{J}_{s,z}\right\vert $ is the spin polarization vector,
which is continuous at the interface to the ferromagnet at $z=0$ and vanishes
at the vacuum interface at $z=d_{N}$. For planar interfaces
\begin{equation}
\boldsymbol{J}_{s,z}(z,\omega)=-\theta_{\mathrm{SH}}J_{c}(\omega
)\mathbf{\hat{y}}-\sigma\partial_{z}\frac{\boldsymbol{\mu}_{s}(z,\omega)}{2e},
\label{isx}%
\end{equation}
where $\theta_{\mathrm{SH}}$ is the spin Hall angle, $\sigma$ the electrical
conductivity, and $J_{c}(\omega)=2\pi J_{c}^{0}\delta(\omega_{a}-\omega)$ the
currents not accounting for spin-orbit interaction. $\boldsymbol{J}%
_{s,z}\left(  d_{N},\omega\right)  =0\ $and$\boldsymbol{J}_{s,z}%
(0,\omega)=\int_{-\infty}^{\infty}\boldsymbol{J}_{s,z}(0,t)e^{-i\omega t}dt$,
where $\boldsymbol{J}_{s,z}\left(  0,t\right)  =\boldsymbol{J}_{s}%
^{\mathrm{T}}+\boldsymbol{J}_{s}^{\mathrm{P}}=\boldsymbol{J}_{s}%
^{\mathrm{(F)}}$ with
\begin{gather}
\boldsymbol{J}_{s}^{\mathrm{T}}=\frac{G_{r}}{e}\hat{\mathbf{M}}\times\left(
\hat{\mathbf{M}}\times\boldsymbol{\mu}_{s}(0)\right)  +\frac{G_{i}}{e}%
\hat{\mathbf{M}}\times\boldsymbol{\mu}_{s}(0),\\
\boldsymbol{J}_{s}^{\mathrm{P}}=\frac{\hbar}{e}\left(  G_{r}\hat{\mathbf{M}%
}\times\partial_{t}\hat{\mathbf{M}}+G_{i}\partial_{t}\hat{\mathbf{M}}\right)
,
\end{gather}
where $\hat{\mathbf{M}}$ is the unit vector along the FI magnetization and
$G^{\uparrow\downarrow}=G_{r}+iG_{i}$ the complex spin-mixing interface
conductance per unit area of the FI$|$N interface. The imaginary part $G_{i}$
can be interpreted as an effective exchange field acting on the spin
accumulation, which is usually much smaller than the real part. A positive
$\boldsymbol{J}_{s}^{\mathrm{(F)}}$ \cite{Jia11,VlietstraX} corresponds here
to up-spins flowing from FI into N. For Pt(Ta) $\omega_{a}\tau_{\mathrm{sf}%
}^{\mathrm{Pt(Ta)}}=1(15)\times10^{-3}$ at the FMR frequency $f_{a}=15.5\,$GHz
with $\tau_{\mathrm{sf}}^{\mathrm{Pt(Ta)}}=0.01(0.15)\,$ps, indicating that
the condition $\omega_{a}\tau_{\mathrm{sf}}^{\mathrm{Pt(Ta)}}\ll1$ is
fulfilled for these metals \cite{Jiao13}. In this limit the frequency
dependence of the spin diffusion length may be disregarded such that
\begin{align}
&  \boldsymbol{\mu}_{s}(z,t)\nonumber\\
&  \rightarrow-\mathbf{\hat{y}}\mu_{s0}(t)\frac{\sinh\frac{2z-d_{N}}{2\lambda
}}{\sinh\frac{d_{N}}{2\lambda}}+\boldsymbol{J}_{s}^{\mathrm{(F)}}%
\frac{2e\lambda}{\sigma}\frac{\cosh\frac{z-d_{N}}{\lambda}}{\sinh\frac{d_{N}%
}{\lambda}},
\end{align}%
\begin{align}
\boldsymbol{J}_{s}^{\mathrm{(F)}}  &  =\frac{\mu_{s0}(t)}{e}\left[
\hat{\mathbf{M}}\times\left(  \hat{\mathbf{M}}\times\mathbf{\hat{y}}\right)
\operatorname{Re}+\hat{\mathbf{M}}\times\mathbf{\hat{y}}\operatorname{Im}%
\right]  T\nonumber\\
&  +\frac{\hbar}{e}\left[  \left(  \hat{\mathbf{M}}\times\partial_{t}%
\hat{\mathbf{M}}\right)  \operatorname{Re}+\partial_{t}\hat{\mathbf{M}%
}\operatorname{Im}\right]  T, \label{jsf}%
\end{align}
where $\mu_{s0}(t)=(2e\lambda/\sigma)\theta_{\mathrm{SH}}J_{c}^{0}\left(
t\right)  \tanh\left[  d_{N}/\left(  2\lambda\right)  \right]  $ with
$J_{c}^{0}(t)=J_{c}^{0}\operatorname{Re}(e^{i\omega_{a}t})$ and $T=\sigma
G^{\uparrow\downarrow}/\left[  \sigma+2\lambda G^{\uparrow\downarrow}%
\coth(d_{N}/\lambda)\right]  $. The ISHE drives a charge current in the
$x$-$y$ plane by the diffusion spin current along $z$. The total charge
current density reads
\begin{equation}
\boldsymbol{J}_{c}(z,t)=J_{c}^{0}(t)\mathbf{\hat{x}}+\sigma\theta
_{\mathrm{SH}}\left(  \nabla\times\frac{\boldsymbol{\mu}_{s}(z,t)}{2e}\right)
.
\end{equation}
The averaged current density over the film thickness is $\overline{J_{c,x}%
}(t)=d_{N}^{-1}\int_{0}^{d_{N}}J_{c,x}(z,t)dz=J_{\mathrm{SMR}}%
(t)+J_{\mathrm{SP}}(t)$ with
\begin{align}
J_{\mathrm{SMR}}(t)  &  =J_{c}^{0}(t)\left[  1-\frac{\Delta\rho_{0}}{\rho
}-\frac{\Delta\rho_{1}}{\rho}\left(  1-\hat{M}_{y}^{2}\right)  \right]
,\label{SMRcurrent}\\
J_{\mathrm{SP}}(t)  &  =J_{r}^{P}\omega_{a}^{-1}\left(  \hat{\mathbf{M}}%
\times\partial_{t}\hat{\mathbf{M}}\right)  _{y}+J_{i}^{P}\omega_{a}%
^{-1}\partial_{t}\hat{M}_{y},\label{ISHEcurrent}\\
&  J_{r(i)}^{P}=\frac{\hbar\omega_{a}}{2ed_{N}\rho}\theta_{\mathrm{SH}%
}\operatorname{Re}\left(  \operatorname{Im}\right)  \eta,\nonumber
\end{align}
where $J_{\mathrm{SMR}}(t)$ and $J_{\mathrm{SP}}(t)$ are SMR rectification and
spin pumping-induced charge currents, $\rho=\sigma^{-1}$ is the resistivity of
the bulk normal metal layer and we recognize the conventional DC\ SMR with
$\Delta\rho_{0}=-\rho\theta_{\mathrm{SH}}^{2}(2\lambda/d_{N})\tanh
(d_{N}/2\lambda)$ and $\Delta\rho_{1}=-\Delta\rho_{0}\operatorname{Re}\eta/2$,
where
\begin{equation}
\eta=\frac{2\lambda\rho G^{\uparrow\downarrow}\tanh\frac{d_{N}}{2\lambda}%
}{1+2\lambda\rho G^{\uparrow\downarrow}\coth\frac{d_{N}}{\lambda}},
\end{equation}
are effective resistivities that do not depend on frequency \cite{Chen}.

ST-FMR experiments employ the AC impedance of the oscillating transverse spin
Hall current caused by the induced magnetization dynamics that is described by
the Landau-Lifshitz-Gilbert (LLG) equation, including the transverse spin
current Eq. (\ref{jsf}),
\begin{equation}
\partial_{t}\hat{\mathbf{M}}=-\gamma\hat{\mathbf{M}}\times
\mathbf{H_{\mathrm{eff}}}+\alpha_{0}\hat{\mathbf{M}}\times\partial_{t}%
\hat{\mathbf{M}}+\frac{\gamma\hbar\boldsymbol{J}_{s}^{(F)}}{2eM_{s}d_{F}},
\end{equation}
where $\mathbf{H_{\mathrm{eff}}}=\mathbf{H_{\mathrm{ex}}}%
+\mathbf{H_{\mathrm{dy}}}$ with an external magnetic field
$\mathbf{H_{\mathrm{ex}}}$ and the sum of the AC current-induced Oersted field
and the (thin film limit of) the dynamic demagnetization $\mathbf{H}%
_{\mathrm{dy}}=\mathbf{H}_{ac}(t)+\mathbf{H}_{d}(t)=\left(  0,H_{ac}%
e^{i\omega_{a}t},-4\pi M_{z}(t)\right)  $. $\gamma$, $\alpha_{0}$, $M_{s}$ and
$d_{F}$ are the gyromagnetic ratio, the Gilbert damping constant of the
isolated film, the saturation magnetization, and the thickness of the FI film, respectively.

We henceforth disregard the very low\ in-plane magnetocrystalline anisotropy
field of $H_{k}\sim3$\thinspace Oe reported \cite{Vlietstra}. The external
magnetic field $\mathbf{H_{\mathrm{ex}}}$ is applied at a polar angle $\theta$
in the $x$-$y$ plane. It is convenient to consider the magnetization dynamics
in the $XYZ$-coordinate system (Fig.~\ref{ACSMR}) in which the magnetization
is stabilized along the $X$-axis by a sufficiently strong external magnetic
field. Denoting the transformation matrix as $R(\theta)$, the magnetization
$\mathbf{M}_{R}(t)=R(\theta)\mathbf{M}(t)$ precesses around the $X$-axis,
where $\mathbf{M}_{R}(t)=\mathbf{M}_{R}^{0}+\mathbf{m}_{R}(t)\approx\left(
M_{s},m_{Y}(t),m_{Z}(t)\right)  $ as shown in Fig. \ref{ACSMR}. $\mathbf{M}%
_{R}^{0}$ and $\mathbf{m}_{\mathrm{R}}(t)$ are the static and the dynamic
components of the magnetization, respectively. The LLG equation in the
$XYZ$-system then becomes $\beta\partial_{t}\mathbf{M}_{R}=-\gamma
\mathbf{M}_{R}\times\mathbf{H}_{\mathrm{eff},R}+\alpha\hat{\mathbf{M}}%
_{R}\times\partial_{t}\mathbf{M}_{R}$ where the effective magnetic field in
the $XYZ$-system is $\mathbf{H}_{\mathrm{eff},R}=H_{X}\hat{\mathbf{X}}%
+H_{Y}e^{i\omega_{a}t}\mathbf{\hat{Y}}+\left(  H_{Z}e^{i\omega_{a}t}-4\pi
m_{Z}(t)\right)  \mathbf{\hat{Z}}$ with $H_{X}=H_{ex}$, $H_{Y}=(H_{ac}%
+H_{i})\cos\theta$ and $H_{Z}=H_{r}\cos\theta$ with
\begin{equation}
H_{r(i)}=\frac{\hbar}{2eM_{s}d_{F}}\theta_{\mathrm{SH}}J_{c}^{0}%
\operatorname{Re}\left(  \operatorname{Im}\right)  \eta,
\end{equation}
a modulated damping $\alpha=\alpha_{0}+\Delta\alpha$ and g-factor
$\beta=1-\Delta\beta$ with $\Delta\alpha\left(  \Delta\beta\right)
=\gamma\hbar^{2}/(2e^{2}M_{s}d_{F})\operatorname{Re}T\left(  \operatorname{Im}%
T\right)  $. For a small-angle precession around the equilibrium direction
$\mathbf{M}_{R}^{0}$, $\mathbf{m}_{R}(t)=(0,\delta m_{Y}e^{i\omega_{a}%
t},\delta m_{Z}e^{i\omega_{a}t})$\ $\left(  \operatorname{Re}[\delta
m_{Y}]\,\operatorname{Re}[\delta m_{Z}]\ll M_{s}\right)  $. Disregarding
higher orders in $\delta m_{Y(Z)}$ in the $R$-transformed LLG equation we
arrive at the (Kittel) relation between AC current frequency and resonant
magnetic field $H_{F}=-2\pi M_{s}+\sqrt{(2\pi M_{s})^{2}+(\omega_{a}%
/\gamma)^{2}}$.

\begin{figure}[ptb]
\includegraphics[width=0.45\textwidth,angle=0]{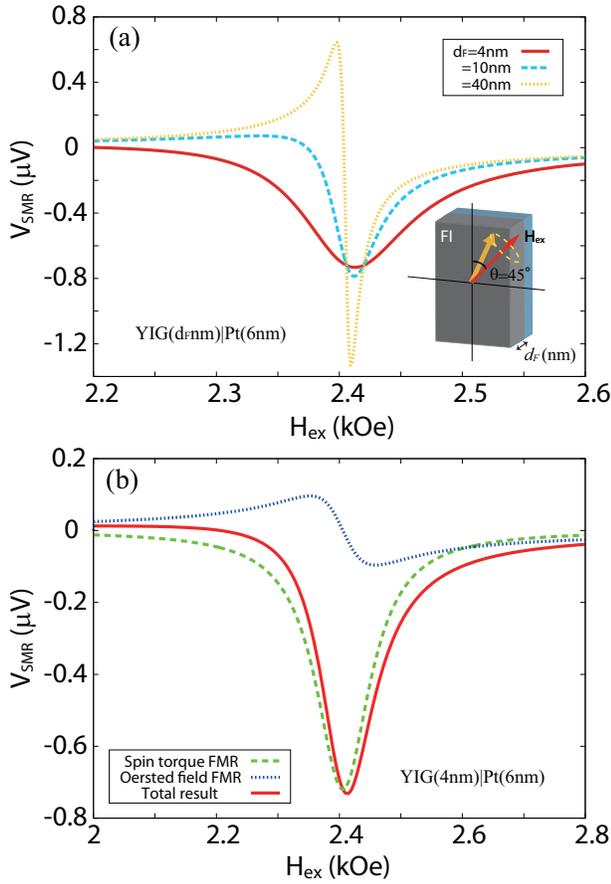}\caption{(a) The
ferromagnet thickness dependence of calculated SMR rectificied voltage for
YIG$|$Pt at $f_{a}=9\,$GHz with current density $J_{c}^{0}=10^{10}%
\ \mathrm{A/m^{2}}$ and F(N) layer length and width $h=w=30\,\mathrm{\mu}$m
and $\theta=45^{\circ}$. (b) $d_{F(N)}=4(6)$ nm.}%
\label{fig.VSMR}%
\end{figure}

\begin{figure}[ptb]
\includegraphics[width=0.45\textwidth,angle=0]{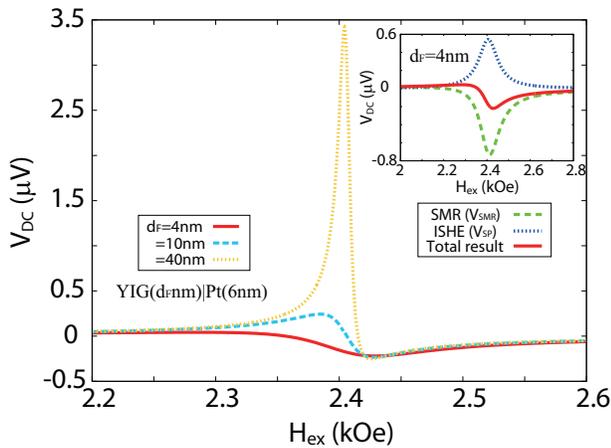}\caption{Dependence
of the ST-FMR spectra on $d_{F}$ at $f_{a}=9\,$GHz and $\theta=45^{\circ}$.
Inset: Contributions by SMR rectification and spin pumping for $d_{F}=4\,$nm.}%
\label{fig.VDC}%
\end{figure}

A DC voltage is generated by two different mechanisms, viz. the time-dependent
oscillations of the SMR in N (spin torque diode effect) and the ISHE generated
by spin pumping. This is quite analogous to electrically detected FMR in which
the magnetization is driven by microwaves in cavities or coplanar wave guides.
In metallic bilayers, the spin pumping signal due to the ISHE can be separated
from effects of the magnetoresistance of the metallic ferromagnet by sample
design and angular dependences \cite{Bai,Obstbaum}. Here we focus on the
current-induced magnetization dynamics that induces down-converted DC and
second harmonic components in the normal metal. Indicating time-average by
$\left\langle \cdots\right\rangle _{t}$ the open-circuit DC voltage is
$V_{DC}=h\rho\langle\overline{J_{c,x}}(t)\rangle_{t}=V_{\mathrm{SMR}%
}+V_{\mathrm{SP}},$ where $V_{\mathrm{X}}=h\rho\langle J_{\mathrm{X}%
}(t)\rangle_{t}$. The SMR rectification and spin pumping induced DC voltage
are \begin{widetext}
\begin{align}
V_{\mathrm{SMR}}&=-\frac{h\Delta\rho_{1}J_{c}^{0}}{4}\frac{F_{S}(H_{\mathrm{ex}}%
)}{\Delta}\left[  C\left(  H_{r}+\alpha H_{ac}\right)  +C_{+}H_{ac}  \frac{H_{\mathrm{ex}}-H_{F}}{\Delta
}\right]  \cos\theta\sin2\theta, \label{VSMR}\\
V_{\mathrm{SP}}&=\frac{h\rho J_{r}^{P}}{4}\frac{F_{S}%
(H_{\mathrm{ex}})}{\Delta}%
C\left[  C_{-}  \frac{H_{r}^{2}+\alpha H_{r}H_{ac}}{\Delta}+C_{+}  \frac{H_{ac}^{2}-\alpha
H_{r}H_{ac}}{\Delta}\right]  \cos\theta\sin2\theta,\label{VSP}
\end{align}
\end{widetext}where $C=\tilde{\omega}_{a}/\sqrt{1+\tilde{\omega}_{a}^{2}}$ and
$C_{\pm}=1\pm1/\sqrt{1+\tilde{\omega}_{a}^{2}}$ with $\tilde{\omega}%
_{a}=\omega_{a}/(2\pi M_{s}\gamma)$, $F_{S}(H_{\mathrm{ex}})=\Delta
^{2}/[\left(  H_{\mathrm{ex}}-H_{F}\right)  ^{2}+\Delta^{2}]$, $\Delta
=\alpha\omega_{a}/\gamma$ the line width, $H_{ac}=2\pi J_{c}^{0}d_{N}/c$ the
Oersted field from the AC current determined by Amp\`{e}re's Law (in the limit
of an extended film), and $c$ speed of light. Using the material parameters
for YIG \cite{Kajiwara} and Pt \cite{Weiler,Obstbaum} shown in
Tables~\ref{tab.YIG} and \ref{tab.Pt} we compute the DC voltages in
Eqs.~(\ref{VSMR}) and (\ref{VSP}). The calculated $V_{\mathrm{SMR}}$ is
plotted in Fig.~\ref{fig.VSMR} as a function of an external magnetic field and
for different $d_{F}$, resolved in terms of the contributions to the FMR
caused by the spin transfer torque (symmetric) and the Oersted magnetic field
(asymmetric). In Fig.~\ref{fig.VDC} we show the total DC voltage with both
spin torque diode and spin pumping contributions. The DC voltage in F$|$Pt
bilayers depends more sensitively on $d_{F}$ for $\mathrm{F=YIG}$ than
$\mathrm{F=Py/CoFeB}$ because spin pumping is more important when the Gilbert
damping is small. ST-FMR measurements are carried out at relatively high
current density, so Joule heating in Pt may cause observable effects, the most
notable being the spin Seebeck effect (SSE), which adds a constant background
DC voltage to the SMR rectification signal \cite{Michael14}.

\begin{table}[ptb]
\begin{ruledtabular}\caption{\label{tab.YIG}Material parameters for the FI layer.}%
\begin{tabular}{ccccc}
& $\gamma\ [\mathrm{T^{-1}s^{-1}}]$ & $M_{s}\ [\mathrm{Am^{-1}}]$ & $\alpha_{0}$\\ \hline
\footnotemark[1]YIG & 1.76$\times 10^{11}$ & 1.56$\times 10^{5}$ & 6.7$\times 10^{-5}$\\
\end{tabular}
\end{ruledtabular}
\footnotemark[1]{Reference~\onlinecite{Kajiwara}.}\end{table}

\begin{table}[ptb]
\begin{ruledtabular}\caption{\label{tab.Pt}Material parameters for the N layer.}%
\begin{tabular}{ccccc}
& $G_{r}\ [\Omega^{-1}\mathrm{m}^{-2}]$ & $\rho\ [\mu\Omega \mathrm{cm}]$ & $\lambda\ [\mathrm{nm}]$
& $\theta_{\mathrm{SH}}$ \\ \hline
Pt & \footnotemark[1]3.8$\times 10^{14}$ & \footnotemark[1]41 & \footnotemark[2]1.4 & \footnotemark[2]0.12 \\
\end{tabular}
\end{ruledtabular}
\footnotemark[1]{Reference~\onlinecite{Weiler},}\footnotemark[2]%
{Reference~\onlinecite{Obstbaum}. }\end{table}

\begin{figure}[ptb]
\includegraphics[width=0.45\textwidth,angle=0]{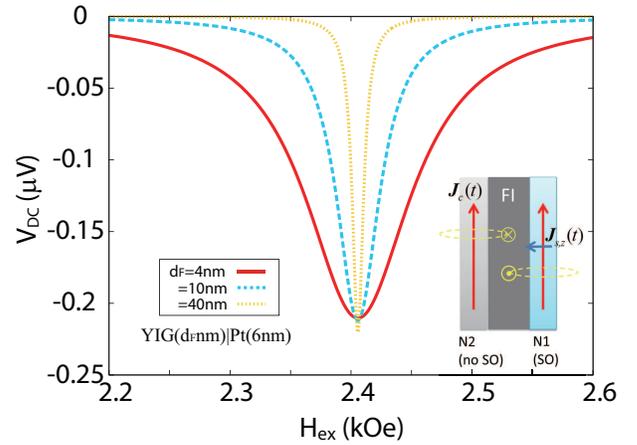}\caption{ST-FMR
spectra dependence on $d_{F}$ in a trilayer set-up to observe the spin torque
induced DC voltages without artifacts of the Oersted field. ($f_{a}=9\,$GHz
and $\theta=45^{\circ}$)}%
\label{fig.Vtri}%
\end{figure}

The ST-FMR spectra in Fig.~\ref{fig.VDC} are enhanced for thicker F layers,
but these are dominated by the Oersted field actuation. These less-interesting
contributions can be eliminated in a tri-layer structure as in
Fig.~\ref{fig.Vtri} in which the magnetic insulator is sandwiched by two
normal metal films with the same electric impedance. The second film N2 should
be Cu or another metal with negligible spin-orbit interaction and thereby
contributions to the ST-FMR, the quality of the YIG%
$\vert$%
N2 interface is therefore less of an issue. In Fig.~\ref{fig.Vtri} we plot
pure ST-FMR signals obtained by setting $H_{ac}=0$ in Eqs.~(\ref{VSMR}) and
(\ref{VSP}), which may now be observed also for thick magnetic layers.

In summary, we predict observable AC current-driven ST-FMR in bilayer systems
consisting of a ferromagnetic insulator such as YIG and a normal metal with
spin-orbit interaction such as Pt. Our main results are the DC voltages caused
by an AC current as a function of in-plane external magnetic field and film
thickness of a magnetic insulator. The DC voltages generated in YIG$|$Pt
bilayers depend sensitively on the ferromagnet layer thickness because of the
small bulk Gilbert damping. The predictions can be tested experimentally by
ST-FMR-like experiments with a magnetic insulator that would yield important
insights into the nature of the conduction electron spin-magnon exchange
interaction and current-induced spin wave excitations at the interface of
metals and magnetic insulators.

This work was supported by KAKENHI (Grants-in-Aid for Scientific Research)
Nos. 22540346, 25247056, 25220910, and 268063, FOM (Stichting voor Fundamenteel
Onderzoek der Materie),\ the ICC-IMR, the EU-RTN Spinicur, EU-FET grant InSpin
612759, and DFG Priority Programme 1538 \textquotedblleft{Spin-Caloric
Transport}\textquotedblright\ (Grant No. BA 2954/1).

\end{document}